\pdfoutput=1
\documentclass[aps,amsmath,preprint,epsf,superscriptaddress,nofootinbib]{revtex4-1}
\usepackage[dvipdfmx]{graphicx}
\usepackage{bm}
\usepackage{color}
\usepackage{amsmath,amssymb}	
\usepackage{hyperref}
\usepackage{setspace}
\usepackage{longtable}
\usepackage{booktabs}
\usepackage{comment}
\usepackage{array}
\usepackage{cancel}
\usepackage{enumitem}
\usepackage{physics}

\def\slashchar#1{\setbox0=\hbox{$#1$} 
\dimen0=\wd0 
\setbox1=\hbox{/} \dimen1=\wd1 
\ifdim\dimen0>\dimen1 
\rlap{\hbox to \dimen0{\hfil/\hfil}} 
#1 
\else 
\rlap{\hbox to \dimen1{\hfil$#1$\hfil}} 
/ 
\fi}

\def\q{\partial}

\def\D{\Delta}

\def\MSb{ {\overline{\rm M}}{\overline {\rm S}  }}

\def\beq{\begin{eqnarray}}
\def\eeq{\end{eqnarray}}

\begin{document}
\newcolumntype{Y}{>{\centering\arraybackslash}p{25pt}} 



\title{The swampland conjecture and the Higgs expectation value}

\author{Koichi Hamaguchi}
\email[e-mail: ]{hama@hep-th.phys.s.u-tokyo.ac.jp}
\affiliation{Department of Physics, Faculty of Science, The University of Tokyo, Bunkyo-ku, Tokyo 113-0033, Japan}
\affiliation{Kavli IPMU (WPI), UTIAS, The University of Tokyo, Kashiwa, Chiba 277-8583, Japan}
\author{Masahiro Ibe}
\email[e-mail: ]{ibe@icrr.u-tokyo.ac.jp}
\affiliation{ICRR, The University of Tokyo, Kashiwa, Chiba 277-8582, Japan}
\affiliation{Kavli IPMU (WPI), UTIAS, The University of Tokyo, Kashiwa, Chiba 277-8583, Japan}
\author{Takeo Moroi}
\email[e-mail: ]{moroi@phys.s.u-tokyo.ac.jp}
\affiliation{Department of Physics, Faculty of Science, The University of Tokyo, Bunkyo-ku, Tokyo 113-0033, Japan}
\affiliation{Kavli IPMU (WPI), UTIAS, The University of Tokyo, Kashiwa, Chiba 277-8583, Japan}

\preprint{UT-18-19}
\preprint{IPMU18-0159}

\date{\today}
\begin{abstract}
  The recently proposed de Sitter swampland conjecture excludes local
  extrema of a scalar potential with a positive energy density in a
  low energy effective theory.  Under the conjecture, the observed
  dark energy cannot be explained by the cosmological constant.  The
  local maximum of the Higgs potential at the symmetric point also
  contradicts with the conjecture.  In order to make the Standard
  Model consistent with the conjecture, it has been proposed to
  introduce a quintessence field, $Q$, which couples to the
  cosmological constant and the local maximum of the Higgs potential.
  In this paper, we show that such a modified Higgs potential
  generically results in a $Q$-dependent Higgs vacuum expectation
  value (VEV).  The $Q$-dependence of the Higgs VEV induces a
  long-range force, which is severely excluded by the tests of the
  equivalence principle.  Besides, as the quintessence field is in
  motion, the Higgs VEV shows a time-dependence, which is also
  severely constrained by the measurements of the time-dependence of
  the proton-to-electron mass ratio.  
 Those constraints require an additional fine-tuning which is justified neither by the swampland conjecture
  nor the anthropic principle. 
  We further show that, even if such an
  unjustified fine-tuning condition is imposed at the tree level,
  radiative corrections upset it. 
  Consequently, we argue that most of the 
  {\it habitable} vacua in the string landscape are in tension with
  the phenomenological constraints.
\end{abstract}
\maketitle

\section{Introduction}
As the string theory is virtually a unique candidate for a theory of quantum gravity, the consistency between a low energy effective field theory
and the string theory is a prime concern for physics beyond the Standard Model.
So far, several consistency conditions have been conjectured from the string theory~\cite{Vafa:2005ui,Ooguri:2006in,Ooguri:2016pdq,Obied:2018sgi}.
Low-energy effective field theories which do not satisfy those conjectures are said to be not in the string landscape,  but in the swampland and are disfavored.

Among various conjectures, the most recent one~\cite{Obied:2018sgi}, 
the so-called de Sitter swampland conjecture, stimulates intensive studies of its phenomenological and cosmological 
consequences~\cite{Vagnozzi:2018jhn,Agrawal:2018own,Andriot:2018wzk,Achucarro:2018vey,Garg:2018reu,Lehners:2018vgi,Kehagias:2018uem,Dias:2018ngv,Denef:2018etk,
Colgain:2018wgk,Brandenberger:2018fdd,Paban:2018ole,Ghalee:2018qeo,Matsui:2018bsy,Ben-Dayan:2018mhe,Chiang:2018jdg,Heisenberg:2018yae,Damian:2018tlf,Kinney:2018nny,Cicoli:2018kdo,
Akrami:2018ylq,Heisenberg:2018rdu,Murayama:2018lie,Marsh:2018kub,Brahma:2018hrd,Choi:2018rze,Das:2018hqy,Danielsson:2018qpa,Wang:2018duq,Brandenberger:2018wbg,
Han:2018yrk,Brandenberger:2018xnf,Matsui:2018xwa}.
Under the conjecture, the scalar potential of a set of scalar fields $\{\phi\}$, $V_{\rm total}$,  satisfies the condition,
\begin{eqnarray}
\label{eq:dSswampland}
M_{\rm PL} | \nabla V_{\rm total} | > c\, V_{\rm total}\ .
\end{eqnarray}
Here, $c$ is a positive constant of ${\cal O}(1)$ and $M_{\rm PL}$ is the reduced Planck scale.
The size of the potential gradient, $|\nabla V_{\rm total}|$, is given by
\begin{eqnarray}
|\nabla V_{\rm total}| = \left(\sum_{\phi} (\q V_{\rm total}/\q\phi)^2\right)^{1/2}\ ,
\end{eqnarray}
for the canonically normalized scalar fields.

Immediate consequences of the conjecture are
\begin{itemize}
\item The observed dark energy cannot be explained by a positive cosmological constant.
\item The local maximum of the Higgs potential at the symmetric point, $H = 0$, is inconsistent with the conjecture.
\end{itemize}
As discussed in Refs.\
\cite{Obied:2018sgi,Denef:2018etk,Murayama:2018lie}, the most
straightforward resolution of these tensions is to couple the
cosmological constant and the Higgs potential to the so-called
quintessence field $Q$, whose coupling is suppressed by the Planck
scale~\cite{Ratra:1987rm, Wetterich:1987fm, Tsujikawa:2013fta}.

In this paper, we show that such a modified Higgs potential
generically predicts a $Q$-dependent vacuum expectation value (VEV) of
the Higgs field.  The $Q$-dependence of the Higgs VEV induces a
long-range force which is severely excluded by the tests of the
equivalence principle~\cite{Wagner:2012ui}.  In addition, the
$Q$-dependence results in a time-dependent Higgs VEV.  We show that
the precise spectroscopic measurements of the proton-to-electron mass
ratio in distant astrophysical systems~\cite[and references
therein]{Martins:2017yxk} put stringent constraint on the time-varying
Higgs VEV.  Consequently, we argue that most of the {\it habitable} vacua in
the string landscape are in tension with the phenomenological
constraints unless there is an additional fine-tuning that is
justified neither by the swampland conjecture nor the anthropic
principle.  We further show that, even if such an unjustified
fine-tuning condition is imposed at the tree level, it is inevitably
violated by radiative corrections.  Therefore, under the de Sitter swampland
conjecture, most of the {\it habitable} vacua in the string landscape
contradict with the observations.

The organization of the paper is as follows.  In Sec.~\ref{sec:Higgs},
we first discuss how we can retrofit the Higgs potential so that it
can be salvaged from the swampland at the tree level argument. We then
show that, such a modified potential generically leads to a
$Q$-dependent Higgs VEV, which is severely constrained by
phenomenological requirements. We argue that an additional fine-tuning
condition is required, which is justified neither by the swampland
conjecture nor the anthropic principle.  In Sec.~\ref{sec:radiative},
we study radiative corrections to the $Q$-dependent Higgs potential,
and show that they generically upset the fine-tuning condition imposed
at the tree level.  The final section is devoted to our conclusions.
In Appendix \ref{sec:qmotion}, we give a rough estimate of the
excursion of the quintessence field from the early universe to the
present.

\section{Salvaging the Higgs potential from swampland}
\label{sec:Higgs}
\subsection{Higgs potential with a tiny dark energy}
The most straightforward way to make the observed dark energy
consistent with the de Sitter swampland conjecture is to introduce the
so-called quintessence field~\cite{Obied:2018sgi}.  Here, we take the
simplest form of the potential of the (real-valued) quintessence
field, $Q$, in the present universe (i.e., in the universe after the
electroweak phase transition):
\begin{eqnarray}
\label{eq:quintess}
 V_Q(Q)  &=& 3\xi_{cc} H_0^2 M_{\rm PL}^2  e^{-c_Q Q/M_{\rm PL }}\ ,
\end{eqnarray}
where $H_0$ is the expansion rate of the present universe, while
$\xi_{cc}$ and $c_Q$ are positive-valued constant parameters.%
\footnote{For $c_Q<0$, we redefine $Q' = - Q$} We set $Q = 0$ as the
present value without loss of generality.  Then, $\xi_{cc}$ is set to
be $\xi_{cc}\simeq\Omega_{DE}\simeq 0.7$~\cite{Aghanim:2018eyx} (with
$\Omega_{DE}$ being the density parameter of the dark energy) to
explain the observed dark energy density (see
Appendix~\ref{sec:qmotion} for more details).  The potential does not
have any local extrema with a positive energy density, and hence it
satisfies the swampland conjecture for $c_{Q} = {\cal O}(1)$.%
\footnote{We assume that the quintessence field satisfies the
  slow-role condition, $c_Q <\sqrt{6}$.}  It should be noted that the
following arguments do not depend on the details of $V_Q(Q)$ as long
as it satisfies the swampland conjecture.

Now, let us discuss how the swampland conjecture restricts the Higgs sector.
The potential for the Higgs field in the Standard Model is given by 
\begin{eqnarray}
\label{eq:Higgs}
V_H(H)  = - M_H^2 |H|^2 + \lambda |H|^4 + \Lambda_{EW}^4 \ ,
\end{eqnarray}
where $M_H^2 > 0$ is a squared Higgs mass parameter and $\lambda > 0$ the
Higgs quartic coupling constant.  A cosmological constant parameter,
$\Lambda_{EW}^4$, is required so that the vacuum energy is cancelled
at the Higgs vacuum expectation value,%
\footnote{Here, $\langle H \rangle$ denotes the VEV of the second
  component of the Higgs doublet.}
\begin{eqnarray}
\langle H\rangle^2 = \frac{v_H^2}{2}  = \frac{M_H^2}{2\lambda}\ ,
\end{eqnarray}
where the fine-tuning condition is
\begin{eqnarray}
\label{eq:FT1}
V_H(\langle H \rangle) = -  \frac{M_H^4}{4\lambda} + \Lambda_{EW}^4 \simeq 0 \ .
\end{eqnarray}

As pointed out in \cite{Denef:2018etk,Murayama:2018lie}, the Higgs potential in Eq.\,(\ref{eq:Higgs}) does not satisfy
the de Sitter swampland conjecture at the symmetric point, $H = 0$.
In fact, the left-hand side of Eq.\,(\ref{eq:dSswampland}),
\begin{eqnarray}
\left. M_{\rm PL}|\nabla V_{\rm total}|\right|_{H=0} = M_{\rm PL} |\q V_{Q}/\q Q| = 
3 c_Q\xi_{cc} H_0^2 M_{\rm PL}^2  e^{-c_Q Q/M_{\rm PL}} \ ,
\end{eqnarray}
is much smaller than the right-hand side,
\begin{eqnarray}
\left. V_{\rm total}\right|_{H=0} \simeq V_H(H=0) = \Lambda_{EW}^4 \simeq 
\frac{M_H^4}{4\lambda} \gg H_0^2 M_{\rm PL}^2 \ .
\end{eqnarray}

An immediate remedy to make the Higgs potential consistent with the de
Sitter swampland conjecture is to retrofit the $\Lambda_{EW}^4$ term to
couple to the quintessence field, i.e.,
\begin{eqnarray}
\label{eq:Higgs2}
V_H^{(a)}(H,Q)  = - M_H^2 |H|^2 + \lambda |H|^4 + \Lambda_{EW}^4e^{-c_H Q/M_{\rm PL}} \ ,
\end{eqnarray}
with $c_H = {\cal O}(1)$.
With the modification, the left-hand side  of Eq.\,(\ref{eq:dSswampland}) becomes
\begin{eqnarray}
\left. M_{\rm PL} |\nabla V_{\rm total}\right|_{H=0} \simeq |\q V_{H}^{(a)}/\q Q|_{H=0} = |c_H| \Lambda_{EW}^4\, e^{-c_H Q/M_{\rm PL}} \ ,
\end{eqnarray}
which is comparable with the right-hand side, 
\begin{eqnarray}
\left. V_{\rm total}\right|_{H=0} \simeq \Lambda_{EW}^4 e^{-c_HQ/M_{\rm PL}}\ .
\end{eqnarray}
In this way, the Higgs potential in Eq.\,(\ref{eq:Higgs2}) can be consistent with the de Sitter swampland conjecture.

The modified Higgs potential in Eq.\,(\ref{eq:Higgs2}), however, has a
serious problem.  At the present vacuum, the fine-tuning condition of
the vacuum energy is given by
\begin{eqnarray}
\label{eq:FT2}
V_H^{(a)}(H=\langle H\rangle,  Q  = 0) = - \left. \frac{M_H^4}{4\lambda} + \Lambda_{EW}^4e^{-c_H Q/M_{\rm PL}}\right|_{Q = 0} \simeq 0 \ .
\end{eqnarray}
However, the quintessence field feels a  strong potential force from the coupling to $\Lambda_{EW}^4$ at the present vacuum,
\begin{eqnarray}
\left. |\q V_{H}^{(a)}/\q Q|\right|_{H=\langle H\rangle} =   |c_H| \frac{\Lambda_{EW}^4}{M_{\rm PL}}\, e^{-c_H Q/M_{\rm PL}}\ ,
\end{eqnarray}
which makes the quintessence field moves from $Q=0$.
Accordingly, the fine-tuning condition in Eq.\,(\ref{eq:FT2}) is immediately violated once the quintessence field evolves in time, within a time scale of $\tau\sim M_{\rm PL}/\Lambda_{EW}^2\sim O(10^{-10})~{\rm sec}$.\footnote{In this case, the quintessence field evolves so rapidly that the observed current Universe is never realized.}
Therefore, although the Higgs potential in Eq.\,(\ref{eq:Higgs2}) is consistent with the swampland conjecture, 
it is not {\it habitable}, and hence, does not satisfy the anthropic principle~\cite{Bousso:2000xa,Susskind:2003kw,Tegmark:2005dy}.

In order to avoid this problem, the Higgs potential needs to be further modified so that the fine-tuning condition of 
the vacuum energy at $H  = \langle H \rangle$ is not affected by the motion of the quintessence field.
Such a requirement can be satisfied, for example, by extending the quintessence-Higgs coupling to
\begin{eqnarray}
\label{eq:Higgs3}
V_H^{(b)}(H, Q) =  - M_H^2e^{-c_M Q/M_{\rm PL}} |H|^2 + \lambda e^{-c_\lambda Q/M_{\rm PL}}|H|^4 + \Lambda_{EW}^4e^{-c_H Q/M_{\rm PL}} \ ,
\end{eqnarray}
where $c_M$ and $c_\lambda$ are ${\cal O}(1)$ coefficients.
For a given value of the quintessence field $Q$, the Higgs VEV is then given by
\begin{eqnarray}
\label{eq:HVEV1}
\left.\langle H \rangle^2\right|_{V^{(b)}_H} = \frac{v_H(X)^2}{2} = \frac{M_H^2 e^{-(c_M -c_\lambda)X}}{2\lambda}\ ,
\end{eqnarray}
where $X=Q/M_{\rm PL}$. As a result, the fine-tuning condition of the vacuum energy is given by
\begin{eqnarray}
\label{eq:FT0}
V_H^{(b)}(\langle H\rangle,Q) = \left(-  \frac{M_H^4}{4\lambda}e^{-(2c_M-c_{\lambda}-c_H)Q/M_{\rm PL}} + \Lambda_{EW}^4\right)e^{-c_H Q/M_{\rm PL}} \simeq 0 \ .
\end{eqnarray}
Therefore, the stability of the small dark energy is achieved by imposing a fine-tuning condition,
\begin{eqnarray}
\label{eq:FT3}
2 c_M - c_{\lambda} - c_H = 0 \ .
\end{eqnarray}

In this way, we arrive at a Higgs potential which is consistent with the de Sitter swampland conjecture and the anthropic principle.
It should be stressed that the additional fine-tuning condition in Eq.\,(\ref{eq:FT3})  for the {\it stability of the small vacuum energy} 
does not make the model less plausible, since  we anyway need to find {\it habitable} vacua in the string landscape~\cite{Bousso:2000xa,Susskind:2003kw,Tegmark:2005dy}.

\subsection{Constraints on the quintessence dependent Higgs VEV}
A crucial feature of the Higgs potential, $V^{(b)}_H$, is that the Higgs VEV generically depends on the quintessence field, as shown in \eqref{eq:HVEV1}. 
This $Q$-dependent Higgs VEV induces effective Yukawa couplings between the quintessence field and the matter fields in the Standard Model,\footnote{The effective Yukawa couplings can also be obtained by diagonalizing the mass matrix of the Higgs and quintessence fields, which leads to a mixing between them in the basis of mass eigenstates.}
\begin{eqnarray}
\label{eq:effF}
{\cal L}_{\rm eff} \simeq \sum_{i=\substack{\rm quarks\\ \rm leptons} }\frac{m_i}{M_{\rm PL}}\left.\frac{d \ln v_H(X)}{d X}\right|_{X=0} Q  \bar{\psi}_{i}\psi_{i}\ ,
\end{eqnarray}
with 
\begin{eqnarray}
\left.\frac{d \ln v_H(X)}{d X}\right|_{X=0} = -\frac{1}{2}(c_M-c_\lambda)\ .
\end{eqnarray}
Here, $m_i$ denotes the mass of the corresponding fermion.

The coupling of the quintessence field to the quarks leads to its
coupling to the nucleons~\cite{Shifman:1978zn},
\begin{eqnarray}
\label{eq:effN}
{\cal L}_{\rm eff} &=& \frac{m_N f_N}{M_{\rm PL}} \left.\frac{d \ln v_H(X)}{d X}\right|_{X=0} Q \bar{\psi_N}{\psi_N}\ ,
\end{eqnarray}
where $m_N$ is the nucleon mass, and $f_N$ is defined as
\begin{eqnarray}
f_N &=& \sum_{q = u,d,s,c,b,t}f_q^N
= \frac{2}{9} + \frac{7}{9}\sum_{q=u,d,s} f^N_q\ ,
\end{eqnarray}
with
\begin{eqnarray}
f_q^N&=&\frac{1}{m_N}\langle N|m_q \bar\psi_{q} \psi_q|N\rangle\ .
\end{eqnarray}
In the following, we use the scalar coupling estimated by using phenomenological and lattice QCD  calculations~\cite{Hoferichter:2017olk},
\begin{eqnarray}
f_N = 0.308(18) \ .
\end{eqnarray}
The isospin violating effect is also estimated to be,
\begin{eqnarray}
f_p-f_n \simeq -1.5\times 10^{-3}\ ,
\end{eqnarray}
with an ${\cal O}(10)$\% accuracy~\cite{Crivellin:2013ipa}.
Altogether, we find that the electron and nucleons couple to the quintessence field,
\begin{eqnarray}
\label{eq:Leff}
{\cal L}_{\rm eff} = q_e Q\bar{\psi}_e {\psi}_e +   q_p Q\bar{\psi}_p {\psi}_p +  q_n Q\bar{\psi}_n {\psi}_n\ ,
\end{eqnarray}
where 
\begin{eqnarray}
  q_e &=& \frac{m_e}{M_{\rm PL}} \left.\frac{d \ln v_H(X)}{d X}\right|_{X=0} \ ,
  \\
  q_{p,n} &=& \frac{m_{p,n} f_{p,n}^N}{M_{\rm PL}} \left.\frac{d \ln v_H(X)}{d X}\right|_{X=0} \ .
\end{eqnarray}

\subsubsection{Long-range force}

The Yukawa interaction of the quintessence field to the electron and the nucleons, Eq.\,(\ref{eq:Leff}), induces a long range force among electrons and nucleons, which are severely constrained by the tests of the equivalence principle~\cite{Wagner:2012ui}. 
In our setup, the parameters $\tilde g$ and $\tilde \psi$ in~\cite{Wagner:2012ui} are identified as
\begin{eqnarray}
\tilde g^2 &=& (q_e+q_p)^2 + q_n^2 \ ,\\
\tan\tilde\psi &=& \frac{q_n}{q_e + q_p}\ .
\end{eqnarray}
The angle $\tilde \psi$ is given by $\tilde \psi \simeq \pi/4$ for
$q_e\ll q_n\simeq q_p$, and the constraint reads
\begin{eqnarray}
\label{eq:long}
\sqrt{(q_e+q_p)^2 + q_n^2} \lesssim 4\times 10^{-24}\ .
\end{eqnarray}
Thus, in the model with the Higgs potential $V_{H}^{(b)}$, we find a stringent constraint,
\begin{eqnarray}
\label{eq:long-range-force-condition}
|c_M- c_\lambda | \lesssim 0.4\times 10^{-4}\ .
\end{eqnarray}
Consequently, the Higgs potential $V_H^{(b)}$ is in tension with observations,
although it is consistent with the de Sitter swampland conjecture and the anthropic principle.

\subsubsection{Time-varying electron-to-proton mass ratio}

Let us also note that the quintessence field is in motion at present
due to the potential force in Eq.\,(\ref{eq:quintess}), which leads to
a non-trivial shift of $Q$.  We have numerically solved the
cosmological evolution of $Q$, and the results are summarized in
Appendix \ref{sec:qmotion}.  For example, the shift of $Q$ from $z=1$
to the present (with $z$ being the redshift parameter) is estimated as
\begin{eqnarray}
\left.\frac{{\mit \D} Q}{M_{\rm PL}}\right|_{z=1} \simeq -0.24 \times c_Q\ .
\end{eqnarray}
Accordingly, the masses of the electron and the proton also depend on
time through the quintessence couplings in Eq.\,(\ref{eq:Leff});
\begin{eqnarray}
\left.\frac{{\mit \D m_e}}{m_e}\right|_{z=1} &=&  \left.\frac{d \ln v_H(X)}{d X}\right|_{X=0}\times  \left.\frac{{\mit \D} Q}{M_{\rm PL}}\right|_{z=1} \ ,\\
\left.\frac{{\mit \D m_p}}{m_p}\right|_{z=1} &=&  f_p\times \left.\frac{d \ln v_H(X)}{d X}\right|_{X=0}\times  \left.\frac{{\mit \D} Q}{M_{\rm PL}}\right|_{z=1}\ .
\end{eqnarray}
Thus, we find that  the ratio of the proton-to-electron mass, $\mu_{pe}=m_p/m_e$, exhibits a time dependence, 
\begin{eqnarray}
\left.\frac{{\mit \Delta}\mu_{pe}}{\mu_{pe}}\right|_{z=1}&\simeq& - 0.7\times   \left.\frac{d \ln v_H(X)}{d X}\right|_{X=0}\times  \left.\frac{{\mit \D} Q}{M_{\rm PL}}\right|_{z=1}\ .
\end{eqnarray}
Such a time dependence of $\mu_{pe}$ is severely constrained by spectroscopic measurements of distant astrophysical systems.
A compilation of the spectroscopic tests~\cite[and references therein]{Martins:2017yxk} amounts to
\begin{eqnarray}
\label{eq:test}
\left.
\frac{{\mit \D}\mu_{pe}}{\mu_{pe}}
\right|_{z<1} = \left(- 0.24\pm  0.09 \right)\times 10^{-6}\ .
\end{eqnarray}
Therefore, we find that the constraint on a time-varying proton-to-electron mass ratio leads to 
\begin{eqnarray}
\label{eq:p-to-e-mass-condition}
c_Q|c_{M}-c_{\lambda}| \lesssim 0.4\times 10^{-5}\ ,
\end{eqnarray}
similarly to the condition obtained from the long-range force
constraint \eqref{eq:long-range-force-condition}.

So far, we assumed that the field which tilts the local maximum of the
Higgs potential (which we denote as $Q_H$ here) and the quintessence
field of the vacuum energy, $Q$, are the same field.  One may argue
that the tensions with the tests of the time-varying
proton-to-electron ratio can be resolved by assuming that these fields
are independent with each other and the motion of the quintessence
field does not affect the Higgs VEV at all even if the Higgs VEV
depends on $Q_H$.  However, as we will discuss, not only the tree
level contribution but also radiative contributions to the vacuum
energy from the Higgs sector are required to be cancelled to explain
the tiny observed dark energy.  Thus, the coupling of $Q_H$ to the
Higgs field inevitably results in a coupling of $Q_H$ to the dark
energy.  Therefore, it is generically expected that $Q_H$ is also in
motion as in the case of $Q$, and hence, the assumption of the
independent $Q_H$ does not resolve the tension, unless there is a
fundamental reason to forbid a time-varying Higgs expectation value.

\subsubsection{Unjustified fine-tuning}

From the phenomenological constraints \eqref{eq:long-range-force-condition} and \eqref{eq:p-to-e-mass-condition}, 
we arrive at an additional fine-tuning condition
\begin{eqnarray}
\label{eq:FT4}
c_M - c_\lambda \simeq 0\ .
\end{eqnarray}
By combining Eqs.\,(\ref{eq:FT0}), (\ref{eq:FT3}) and (\ref{eq:FT4}),
the Higgs potential is then restricted to a form 
\begin{eqnarray}
  \label{eq:Higgs4}
  V_H^{(c)}(H, Q) = 
  \left(- M_H^2 |H|^2 + \lambda |H|^4 + \Lambda_{EW}^4\right)e^{-c_H Q/M_{\rm PL}} 
  = 
  \lambda e^{-c_H Q/M_{\rm PL}} \left(|H|^2 - \frac{1}{2} v_{H}^2\right)^2\ ,
  \nonumber \\
  \label{V^(c)}
\end{eqnarray}
which corresponds to the Higgs-quintessence coupling proposed in
\cite{Denef:2018etk}.\footnote{See Ref.~\cite{Cicoli:2018kdo} for related discussions on the  Higgs-quintessence couplings.
}  As the quintessence coupling is factored out as
an overall factor, the Higgs VEV does not depend on the quintessence
field. 

It should be stressed that the additional fine-tuning in
Eq.~\eqref{eq:FT4}, or more generically an independence of the Higgs
VEV on the quintessence field, is justified neither by the swampland
conjecture nor the anthropic principle, but it is required from purely
phenomenological reasons.%
\footnote{As discussed in \cite{Hall:2014dfa}, the Higgs VEV in the
  era of the Big-Bang Nucleosynthesis (BBN) is allowed to be different
  from the current value by a factor ${\cal O}(1)$ for habitable
  universe.  See also \cite{Harnik:2006vj} for a related discussion.}
In other words, under the assumption that the swampland conjecture is
satisfied by a quintessence-like field,  most of the {\it habitable} vacua
in the string landscape are excluded by the observational constraints,
unless there is an additional fine-tuning that is justified neither by
the conjecture nor the anthropic argument.

In principle, it is possible to assume $Q$-dependent Yukawa couplings 
which make the fermion masses independent of the quintessence field.
Again, however, such fine-tunings are not required by the swampland conjecture nor the anthropic principle.
Therefore, this possibility does not explain why there is no long-range force nor why the proton-to-electron mass ratio is time-independent.
Besides, the $Q$-dependent Yukawa couplings lead to $Q$-dependences of the gauge couplings through radiative corrections, which are also restricted by the tests on the time-varying coupling constants~\cite[and references therein]{Martins:2017yxk}.
In this paper, we do not pursue these possibilities any further.

\section{Radiatively induced $Q$-dependent Higgs VEV}
\label{sec:radiative}
As we have discussed in the previous section, it is required to choose Higgs-quintessence couplings
so that the Higgs VEV does not depend on the quintessence field,
 although such a condition is not required from the de Sitter swampland conjecture nor the anthropic principle.
The Higgs potential $V_H^{(c)}$ in Eq.\,(\ref{eq:Higgs4}) is the simplest example which satisfies this condition.
In this section, we assume $V_H^{(c)}$ and discuss whether the Higgs VEV remains independent of the quintessence field 
when we consider radiative corrections in the low energy effective field theory.

\subsection{Wilsonian approach}   
To obtain a rough idea how the radiative corrections affect the low
energy effective field theory, let us first assume that the Higgs
potential in a Wilsonian effective action at around the Planck scale
is given by $V_{H}^{(c)}$.  We also assume that the other couplings
such as the gauge coupling constants and the Yukawa coupling constants
do not depend on the quintessence field.  
This assumption is motivated
by the fact that the $Q$-dependences of them are severely constrained
by the tests of the equivalence principle, by the test of the
time-variation of the fundamental couplings, and by the BBN
constraints~\cite{Campbell:1994bf,Wagner:2012ui,Martins:2017yxk}.

In this setup, the squared Higgs mass parameter 
and the quartic coupling and 
at a low energy scale receives radiative corrections
\begin{eqnarray}
  M_H^2(\mu_R) &=& 
  M_H^2 e^{-c_HQ/M_{\rm PL}}
  \left(
    1 + \int^{\mu_R}_{M_{\rm PL}} \frac{d{\mu_R'}}{\mu_R'} \gamma_{M_H^2}(\mu_R') 
  \right)
  \simeq M_H^2(\mu_R)|_{Q=0} \, e^{-c_HQ/M_{\rm PL}} \ ,
  \label{eq:1LM}
\end{eqnarray}
and
\begin{eqnarray}
  \lambda(\mu_R) &=& \lambda e^{-c_HQ/M_{\rm PL}}  + 
  \int^{\mu_R}_{M_{\rm PL}} \frac{d{\mu_R'}}{\mu_R'}\beta_\lambda(\mu_R')
  \nonumber \\
  &\simeq& 
  \left[ \lambda(\mu_R)|_{Q=0}\,+ \left(e^{c_HQ/M_{\rm PL}}-1\right)
    \int^{\mu_R}_{M_{\rm PL}} \frac{d{\mu_R'}}{\mu_R'}\beta_\lambda(\mu_R')\right]
  \, e^{-c_HQ/M_{\rm PL}} \ , 
  \label{eq:1LLAM}
\end{eqnarray}
where $\mu_R^{(\prime)}$ denotes the renormalization scale,
$\beta_{\lambda}$ the beta functions of $\lambda$, and
$\gamma_{M_H^2}$ the anomalous dimension of $M_H^2$.  In Eqs.\,(\ref{eq:1LM}) and (\ref{eq:1LLAM}), we rewrite
them by using the low-energy parameters for $Q = 0$.  Here, we
neglected the $Q$-dependences of $\beta_{\lambda}$ and $\gamma_{M_H^2}$ as they are
dominated by $Q$-independent interactions, i.e, the top Yukawa
and the gauge interactions.

Once the low energy parameters are given by Eqs.\,(\ref{eq:1LM}) and
(\ref{eq:1LLAM}), it is no longer possible to factor out the
quintessence couplings from the Higgs potential. Thus, the Higgs VEV
has a non-trivial dependence on the quintessence field; for $Q\ll
M_{\rm PL}$, the Higgs VEV is obtained as
\begin{eqnarray}
  \langle H \rangle^2 &\simeq& 
  \left.\frac{M_H^2(\mu_R)}{2\lambda(\mu_R)}\right|_{Q=0}
  \times 
  \left(
    1- \frac{c_H Q/M_{\rm PL}}{\lambda(\mu_R)|_{Q=0}}\times 
    \left.\int \frac{d{\mu_R'}}{\mu_R'}\beta_\lambda(\mu_R')\right|_{Q=0}
  \right) \ .
\end{eqnarray}
Thus, the shift of the quintessence field induces the shift of
the VEV.  
Numerically, we found 
\begin{eqnarray}
  \label{eq:DV}
  \left.\frac{d \ln v_H(X)}{d X}\right|_{X=0}
  \simeq -0.5 \times c_H \ , 
\end{eqnarray}
where $\lambda(m_t) \simeq 0.126$ and the integration of $\beta_\lambda$ is 
\begin{eqnarray}
  \int_{M_{\rm PL}}^{m_t} \frac{d{\mu_R'}}{\mu_R'}\beta_{\lambda}(\mu_R')  \simeq 0.13\ ,
\end{eqnarray}
obtained by using {\tt RGErun2}.
Here, we take the  top quark mass, $m_t = 173$\,GeV, as the low energy renormalization scale.%
\footnote{The relative error of the integration of Eq.\,(\ref{eq:DV}) is of ${\cal O}(10^{-1})$ 
which is dominated by the choice of $\mu_R$.  
We will  provide more precise discussion including the renormalization
  conditions in the next subsection.}

Thus, from the constraints on the long range force in Eq.\,(\ref{eq:long}), we again obtain an upper limit,
\begin{eqnarray}
\label{eq:cHinW}
|c_H| \lesssim 0.4\times 10^{-4}\ ,
\end{eqnarray}
while the tests of the time-variation of the proton-to-electron mass ratio in Eq.\,(\ref{eq:test}) gives
\begin{eqnarray}
\label{eq:cHcQinW}
c_Q|c_H| \lesssim 0.4\times 10^{-5}\ .
\end{eqnarray}
Therefore, we find that the Higgs potential $V_H^{(c)}$ is in tension
with tests of the equivalence principle and the time-variation of the
proton-to-electron mass ratio.

In the above argument, we have implicitly assumed that the quadratic
and the quartic divergences appearing in the Higgs squared mass
parameter and in the cosmological constant term are fine-tuned by
local counter terms even if they have non-trivial dependences on the
quintessence field.  This means that the Higgs potential at high
energy scale includes local terms of the form:
\begin{eqnarray}
V_H =  V^{{c}}_H + \Lambda_2^2(X) |H|^2 + \Lambda_{4}^4(X)\ , 
\end{eqnarray}
in addition to $V_{H}^{(c)}$.  Here, $\Lambda_2^2(X)$ and
$\Lambda_4^4(X)$ are functions of the quintessence field which are
introduced to cancel the $Q$-dependent quadratic and quartic
divergences.  These assumptions are justified by the anthropic
principle, since otherwise, the vacuum is no more
habitable~~\cite{Bousso:2000xa,Susskind:2003kw,Tegmark:2005dy}.

We comment here that, as the local term $\Lambda_2^2(X)|H|^2$ is added
to the high energy Lagrangian, Eq.\,(\ref{eq:1LM}) does not hold in
general.  With such a term, there should exist a residual $Q$-dependent mass term due to the $\Lambda_2^2(X)|H|^2$ term which
is not proportional to $e^{-c_H Q/M_{\rm PL}}$ in general.  Such an
observation does not weaken the constraints in Eqs.\,(\ref{eq:cHinW})
and (\ref{eq:cHcQinW}).

As another comment, it is also tempting to ask whether it is possible
to realize a phenomenologically viable scenario by assuming low energy
parameters giving rise to the Higgs potential of the form of
$V_{H}^{(c)}$ (see Eq.~\eqref{V^(c)}), i.e.,
\begin{eqnarray}
  \label{eq:MH2}
  M_H^2(\mu_R\simeq m_t ) &\simeq& 
  M_H^2(\mu_R \simeq m_t)|_{X=0} e^{-c_H Q/M_{\rm PL}}\ , \\
  \label{eq:lambda}
  \lambda(\mu_R\simeq m_t ) &\simeq& 
  \lambda(\mu_R \simeq m_t)|_{X=0} e^{-c_H Q/M_{\rm PL}}\ .
\end{eqnarray}
These renormalization conditions can be satisfied by adding local
counter terms of the quartic coupling to $V_{H}^{(c)}$ at the high
energy scale with appropriate quintessence field dependences.  As we
will see shortly, however, the Higgs VEV still depends on quintessence
field even if the renormalization conditions in Eqs.\,(\ref{eq:MH2})
and (\ref{eq:lambda}) are imposed.  Let us also stress again that such
fine-tuning conditions are not supported by the de Sitter swampland
conjecture nor by the anthropic principle.


\subsection{Analysis in the $\MSb$ prescription}
To make our statement more precise and concrete, let us consider an effective field theory where the bare Higgs potential is given by
\begin{eqnarray}
  \label{eq:Higgs5}
  V_{HB}(H_B)  = 
  - M_{HB}^2(X) |H_B|^2 + \lambda_B(X) |H_B|^4 + \Lambda_{EWB}^4(X) \ ,
\end{eqnarray}
with the subscript $B$ denoting the bare parameters and fields.  The
$Q$-dependences of the coefficient parameter functions
are not specified at this point.  As for the other coupling constants
in the Standard Model, we assume that they are independent of the
quintessence field.  We treat the quintessence field as a background
field and do not consider the path-integration of the quintessence
field.

In this setup, the quantum effective potential of the Higgs boson is given by
\begin{eqnarray}
\label{eq:Veff}
V_{H\rm eff}(H) = -M_H^{2(\MSb)}(X) |H|^2 
+ \lambda^{(\MSb)}(X) |H|^4 + \Lambda_{EW}^{4(\MSb)}(X)  +V_{H\rm eff}^{(1)} + \cdots\ ,
\end{eqnarray}
where the first three terms are renormalized tree-level contributions
while $V_{H\rm eff}^{(n)}$ $(n>0)$ are $n$-loop contributions.%
\footnote{In this paper, the quantum effective potential denotes the
  one calculated perturbatively for a quantum state whose wave
  functional is localized at around a particular field value.  It differs
  from the one defined by the Legendre transformation of the partition
  function of the connected Green functions,
  $W[J]$~\cite{Weinberg:1987vp}.}  At the tree level, i.e., neglecting
$V_{H\rm eff}^{(n)}$ with $n>0$, the Higgs VEV is given by
\begin{eqnarray}
  \label{eq:tree}
  \left.v_H^{(\MSb)2}(X)\right|_{\rm tree}=
  \frac{M_H^{2(\MSb)}(X)}{\lambda^{(\MSb)}(X)}\ .
\end{eqnarray}
In the following, we discuss how the Higgs VEV behaves after taking
into account radiative corrections.

We adopt the $\MSb$ prescription for the renormalization for a given
value of $X$.  The one-loop effective potential $V^{(1)}_{H\rm eff}$
is given by~\cite{Ford:1992pn,Martin:2013gka}
\begin{eqnarray}
\label{eq:Veff1}
16\pi^2 V^{(1)}_{H\rm eff} &= &
\frac{F_H^2}{4}
\left(
\overline\ln\, F_H - \frac{3}{2}
\right)
+ 
\frac{3 F_G^2}{4}
\left(
\overline\ln\, F_G - \frac{3}{2}
\right)
-3 F_T^2
\left(
\overline\ln\,F_T - \frac{3}{2}
\right)
\cr
&&+\frac{3 F_W^2}{2}
\left(
\overline\ln\, F_W - \frac{5}{6}
\right)
+\frac{3 F_Z^2}{4}
\left(
\overline\ln\, F_Z - \frac{5}{6}
\right)\ ,
\end{eqnarray}
in the $\MSb$ prescription. 
The functions $F$'s are given by
\begin{eqnarray}
  F_H &=& - M_H^{2(\MSb)}(X)  + 3 \lambda^{(\MSb)}(X) h^2 \ , \\
  F_G &=&- M_H^{2(\MSb)}(X)  +  \lambda^{(\MSb)}(X) h^2 \ ,\\
  F_T &=& y_t^{(\MSb)\,2} h^2 /2 \ , \\
  F_W &=& g^{(\MSb)\,2} h^2/4  \ ,\\
  F_Z &=& (g^{(\MSb)\,2}+g^{\prime(\MSb)\,2} ) h^2/4 \ , \\
  \overline \ln F &=& \ln F/\mu_R^2 \ ,
\end{eqnarray}
where $y_t$, $g$ and $g'$ are the top Yukawa and the gauge coupling
constants of the $SU(2)_L$ and $U(1)_Y$ gauge interactions in the
$\MSb$ prescription, respectively.  Here, we parametrize the Higgs
doublet as follows without loss of generality,
\begin{eqnarray}
  H = \frac{1}{\sqrt{2}}(0, h)^T \ .
\end{eqnarray}

Following Refs.\,\cite{Ford:1992pn,Martin:2013gka}, we define the
$\MSb$ VEV, $v_H^{(\MSb)}(X)$, by the field value of $ h$ which
minimizes $V_{H\rm eff}$ in the $\MSb$ prescription.  Then, the $\MSb$
Higgs VEV does not satisfy the tree-level relation given in Eq.\
\eqref{eq:tree}.  As the VEV of the renormalized Higgs field is not a
physical observable, its definition beyond the tree level is arbitrary
and a matter of convention.  The advantage of the definition in
Refs.\,\cite{Ford:1992pn,Martin:2013gka} is that the Higgs tadpole
diagrams are cancelled by definition.  With the present definition of
$v_H^{(\MSb)}(X)$, the pole electron mass is given by
\begin{eqnarray}
m_e(X) = \frac{1}{\sqrt 2} y_e^{(\MSb)} v_H^{(\MSb)}(X) \times \left( 1 - \Re\Sigma_{S}(m_e^2)- \Re\Sigma_{V}(m_e^2)
\right)\ ,
\end{eqnarray}
where $\Sigma_{V,S}(p^2)$ are defined by the free electron self-energy, $\Sigma_e$, 
\begin{eqnarray}
\Sigma_e(p) = i {\slashchar p} \,\Sigma_V(p^2)  + i\, m_e\, \Sigma_S(p^2)    \ ,
\end{eqnarray}
at one loop.%
\footnote{Since the electron mass is much smaller than those of $Z$
  and $W$ bosons appearing in $\Sigma_e$, it is not practical to
  calculate $m_{e}$ by using $\Sigma_e$ obtained in the
  Standard Model.  Rather, we need to match the $\MSb$ electron
  masses in the Standard Model and in the low energy effective theory
  below the electroweak scale.  Those procedures do not affect our
  argument, though.  }  As the Higgs tadpole diagrams automatically vanish
  in $\Sigma_{e}$, the $Q$-dependence of $m_e(X) $ is dominated by the one through
$v_H^{(\MSb)}(X)$.  The $Q$-dependence though
$M_H^{2(\MSb)}(X)$ comes from the Higgs-electron loop contribution to
$\Sigma_{S,V}(m_e^2)$, which is proportional to the electron Yukawa
coupling squared, and is numerically unimportant.  As a result, we
find,
\begin{eqnarray}
  \left.\frac{d \ln m_e(X)}{d X}\right|_{X=0} \simeq 
  \left.\frac{d \ln v_H^{(\MSb)}(X)}{d X}\right|_{X=0}\ .
\end{eqnarray}
Similarly, the quark masses also depend on the quintessence field via
$v_{H}^{(\MSb)}$, and we find
\begin{eqnarray}
\label{eq:Ncouple}
  \left.\frac{d \ln m_N(X)}{d X}\right|_{X=0} \simeq
  f_N\left.\frac{d \ln v_H^{(\MSb)}(X)}{d X}\right|_{X=0}\ .
\end{eqnarray}
It should be noted that the quintessence field couples to the top quark
not only through $v_{H}^{(\MSb)}(X)$ but also through radiative corrections in which the 
Higgs boson is circulating.
For now, we neglect these couplings and we will come back to this point later.

Now, let us discuss renormalization conditions.  From a perspective of
low energy effective field theory, we only know the Higgs potential
parameters in the present universe, i.e. $X = 0$.  For $X\neq 0$,
there is no experimental data to determine the renormalization
conditions.  Thus, we may, for example, {\it impose} $X$-dependences
of the parameter functions of $X$, so that
\begin{eqnarray}
  \label{eq:LQ1}
  M_H^{2(\MSb)}(X) &=& M_H^{2(\MSb)}(X = 0) e^{-c_H X} \ , \\ 
  \label{eq:LQ2}
  \lambda^{(\MSb)}(X) &=& \lambda^{(\MSb)}(X= 0) e^{-c_H X} \ ,
\end{eqnarray}
at a low energy scale such as the pole mass of the top quark, $\mu_R =
m_t^{(\rm pole)}$.
These conditions correspond to the ones in
Eqs.\,(\ref{eq:MH2}) and (\ref{eq:lambda}) in our argument in the
Wilsonian approach.  With these conditions, the tree-level Higgs VEV,
$v_{H}^{(\MSb)}|_{\rm tree}$, is independent of the quintessence
field.
We call these renormalization conditions as the low energy
quintessence (LQ) prescription.  It should be stressed here that
conspiratorial fine-tunings are hidden in the renormalization
conditions in Eqs.\,(\ref{eq:LQ1}) and (\ref{eq:LQ2}) from a
perspective of the high-energy theory.

Beyond the tree level, the Higgs VEV does not satisfy the tree-level
relation.  At the one-loop level, for example, the shift of the VEV is
roughly given by
\begin{eqnarray}
  \label{eq:Vshift0}
  \frac{
    v_{H}^{(\MSb)}(X) - 
    \left.v_{H}^{(\MSb)}\right|_{\rm tree}
  }{\left.v_{H}^{(\MSb)}\right|_{\rm tree}
  }
  \simeq 
  \frac{1}{2 \lambda^{(\MSb)}(X) \left.v_{H}^{(\MSb)3}\right|_{\rm tree}}
  \frac{\partial V^{(1)}}{\partial h} 
  \sim - \frac{9}{64\pi^2\lambda^{(\MSb)}(X)} y_t^{(\MSb)\, 4}\ ,
\end{eqnarray}
where we keep only the top Yukawa contribution in Eq.\,(\ref{eq:Veff1}) for presentation purpose.
Thus, the deviation of the Higgs VEV from the tree-level relation in Eq.\,(\ref{eq:tree}) results in
a non-trivial  quintessence field dependence, which is enhanced by $1/\lambda^{(\MSb)}$.

Our numerical analysis is as follows.
We first
calculate $v_{H}^{(\MSb)}(X=0)$ at $\mu_R = m_t^{(\rm pole)}$ from the
relation between $v_{H}^{(\MSb)}(X=0)$ and the Fermi coupling constant
$G_\mu$ in \cite{Degrassi:2012ry}, which leads to%
\footnote{A nominal uncertainty, $v_{H}^{(\MSb)}(X=0) = 246.7711\pm
  0.0015 $ ($\mu_R = m_{\rm top}^{(\rm pole)}$), is dominated by the
  error of the top quark mass, $m_t^{(\rm pole)} = 173.0 \pm
  0.4$\,GeV~\cite{Tanabashi:2018oca}, although we do not need a very
  precise value of it.  }
\begin{eqnarray}
v_{H}^{(\MSb)}(X=0) = 246.8\,{\rm GeV} \ .
\end{eqnarray}
Then, we obtain the quartic coupling and the Higgs mass parameters at $X=0$, 
\begin{eqnarray}
  \label{eq:MSBlam}
  \lambda_H^{(\MSb)}(X=0) &=&  0.1261\pm 0.0003\ , \\
  M_H^{2(\MSb)}(X=0) &=&  (92.9\pm0.1\,{\rm GeV})^2\ , 
  \label{eq:MSMH}
\end{eqnarray}
at $\mu_R = m_t^{(\rm pole)}$.  
In evaluating these values, we use {\tt SMH}~\cite{Martin:2014cxa}, which
takes full two-loop and leading three-loop corrections into account.%
\footnote{In {\tt SMH}, the loop integrations are handled by {\tt
    TSIL}~\cite{Martin:2005qm}. }  
The uncertainties quoted here do not include the ones from the choice of the renormalization scale.
The input parameters for {\tt SMH}
are taken to be
\begin{eqnarray}
m_{Z}^{(\rm pole)} &=&  91.1876 \pm 0.0021\, {\rm GeV} \ , \\
m_{W}^{(\rm pole)} &=& 80.379 \pm 0.012 {\rm GeV} \ , \\
m_{h}^{(\rm pole)} &=& 125.18 \pm 0.16 {\rm GeV}  \ , \\
m_t^{(\rm pole)} &=& 173.0 \pm 0.4  {\rm GeV} \ ,\\
\alpha_s^{(\MSb)}(m_{z}) &=& 0.1181\pm 0.0011 \ ,
\end{eqnarray}
(see Ref.~\cite{Tanabashi:2018oca}).  The uncertainties of the
parameters in Eqs.\,(\ref{eq:MSBlam}) and (\ref{eq:MSMH}) are
dominated by the uncertainty of $m_{h}^{(\rm pole)}$.  
  The values $v_H^{(\MSb)}(X=0)$,
$\lambda_H^{(\MSb)}(X=0)$, and $M_H^{2(\MSb)}(X=0)$ are taken as
reference values to estimate the shift of the Higgs VEV for $X\neq 0$.
As we are interested in $d\ln v_{H}^{(\MSb)}/dX$, the uncertainties of
those parameters are cancelled at the leading order.

\begin{figure}[t]
\begin{center}
  \includegraphics[width=.5\linewidth]{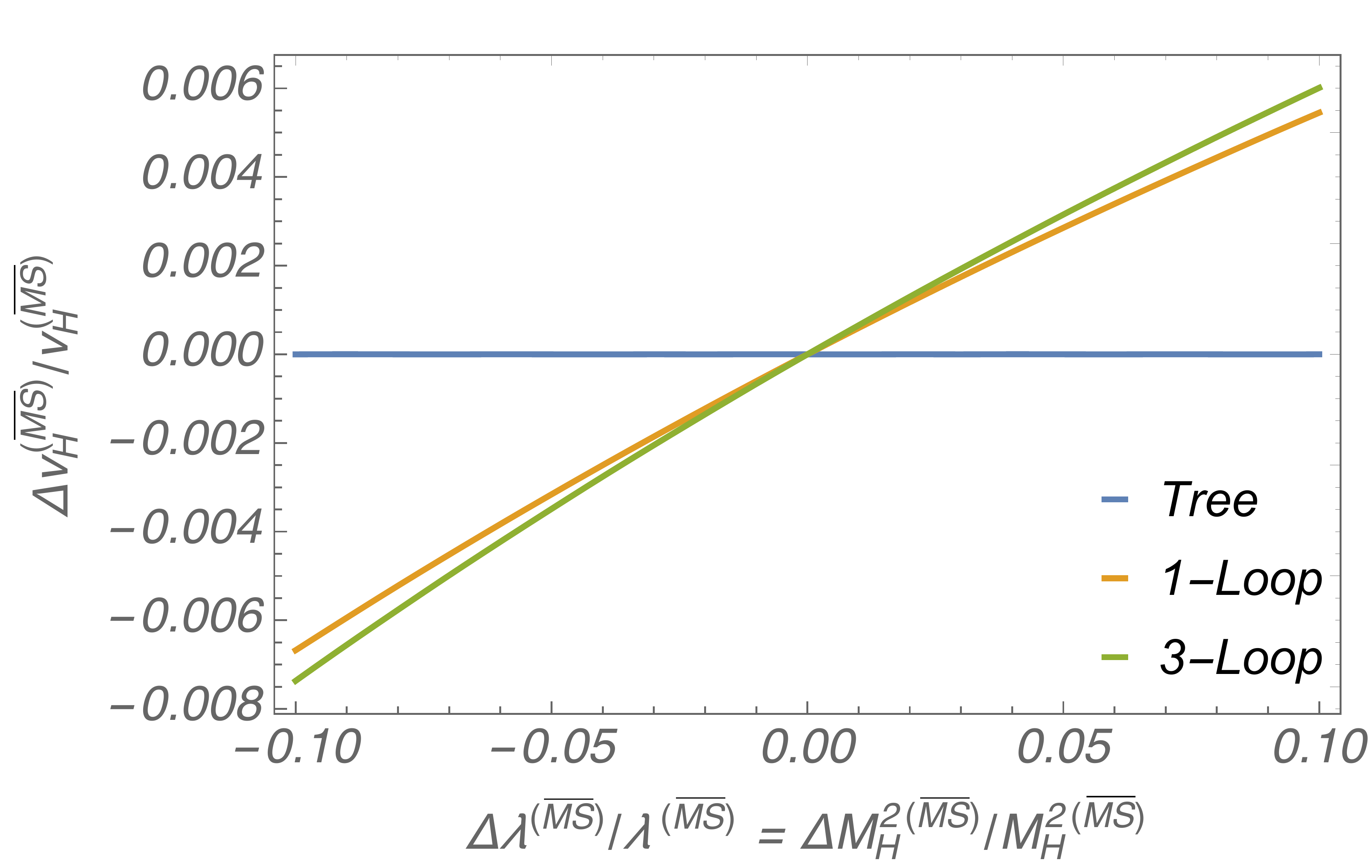}
  \end{center}
\caption{\sl \small  
The shift of the vacuum expectation value of the Higgs VEV in the LQ prescription in Eq.\,(\ref{eq:LQ1}) and (\ref{eq:LQ2}),
where {\tt SMH} is used  for a numerical calculation at tree, one-loop, and three-loop level. 
The figure shows that the Higgs VEV does not shift at the tree-level due to the tree level relation of the Higgs VEV in Eq.\,(\ref{eq:tree}).
Beyond the tree-level, the Higgs VEV is shifted by changing $\lambda^{(\MSb)}$.
}
\label{fig:dv}
\end{figure}

In Fig.\,\ref{fig:dv}, we show how $v_H^{(\MSb)}$ shifts in the LQ
prescription, where $\lambda^{(\MSb)}$ and $M_H^{2(\MSb)}$ are changed
while $M_H^{2(\MSb)}/\lambda^{\MSb}$ is fixed.  Here, we again utilize
{\tt SMH}~\cite{Martin:2014cxa} to obtain $v_H^{(\MSb)}(X\neq 0)$ for
given $\lambda^{(\MSb)}(X\neq 0)$ and $M_H^{2(\MSb)}(X\neq 0)$.  The
figure shows that, based on a calculation including the leading three-loop
effects, the $\MSb$ Higgs VEV changes as%
\footnote{We choose the renormalization scale to be $\mu_R = m_{\rm
    t}^{(\rm pole)}$ even for $X \neq 0$.  }
\begin{eqnarray}
\label{eq:Vshift}
\frac{{\mit \D}v_{H}^{(\MSb)}}{v_H^{(\MSb)}} \simeq 0.07\times 
\frac{{\mit \D}\lambda^{(\MSb)}}{\lambda^{(\MSb)}} \ .
\end{eqnarray}
The rather large $X$ dependence in Eq.\,(\ref{eq:Vshift}) stems from the fact that
the shift of the VEV from the tree-level relation is enhanced by $1/\lambda^{(\MSb)}$ (see Eq.\,(\ref{eq:Vshift0})).
Combining with the renormalization conditions of the LQ prescription
(Eqs.\,(\ref{eq:LQ1}) and (\ref{eq:LQ2})), we find
 \begin{eqnarray}
 \label{eq:HVEV3}
\left.\frac{d \ln v^{(\MSb)}_H(X)}{d X}\right|_{X=0}
\simeq -0.07 \times c_H\ . 
\end{eqnarray}
As a result, we find that the tests of the equivalence principle and the time-variation of the proton-to-electron mass ratio leads to 
slightly weakened conditions,
\begin{eqnarray}
c_H \lesssim 0.3\times 10^{-3}\ ,
\end{eqnarray}
and 
\begin{eqnarray}
c_Q|c_H| \lesssim 0.3\times 10^{-4}\ ,
\end{eqnarray}
respectively.%
\footnote{Recently, it has been argued that the change of the Higgs quartic coupling by about a ten percent level at low energy due to the quintessence field
may stabilize the Higgs vacuum in the Standard Model~\cite{Han:2018yrk}. 
Our result shows that such a possibility has a tension with the constraint in Eq.\,(\ref{eq:test}).}
Thus, the Higgs potential $V_H^{(c)}$ with ${\cal O}(1)$ coefficients is in tension with the observational constraints
even in the LQ prescription. 

So far, we have imposed the renormalization conditions (\ref{eq:LQ1})
and (\ref{eq:LQ2}) at the electroweak scale without assuming any
particular high energy theory.  As another possibility, we may
impose them 
at $\mu_R \simeq M_{\rm PL}$.  This possibility corresponds to the
case discussed in the previous subsection, i.e., the case with the
Wilsonian effective action with $V_H^{(c)}$ at around the Planck
scale.  We call these renormalization conditions as the high energy
$\MSb$ quintessence (HQ) prescription.  In this case, $d \ln v_H^{(\MSb)}(X)/d
X|_{X=0}\simeq -0.5\times c_H$ (see Eq.~\eqref{eq:DV}), which is an order
of magnitude larger than the LQ case.  Therefore, we again conclude
that the HQ prescription defined in the $\MSb$ scheme is in a strong
tension with the current constraints (see Eqs.\,(\ref{eq:cHinW}) and (\ref{eq:cHcQinW})).

Several comments are in order.  In the LQ prescription, the reason why
$v_{H}^{(\MSb)}$ shifts is that the effective Higgs potential is
modified by the $n$-loop contributions, $V_{H\rm eff}^{(n)}$.  At the
one-loop level, for example, the most relevant terms for the VEV shift
are the second terms of each contribution in Eq.\,(\ref{eq:Veff1}).
As they are proportional to the tree-level terms in Eq.\,(\ref{eq:Veff}),
we can eliminate the effects of those times for $X\neq 0$ by carefully adjusting the renormalization conditions 
such that,
\begin{eqnarray}
\label{eq:awful1}
M_H^{2(\MSb)}(X)
&=&
M_H^{2(\MSb)}(X=0) e^{-c_H X} 
+(1-e^{-c_H X})  \frac{9}{16\pi^2} \lambda^{(\MSb)} m_H^{2(\MSb)} + \cdots\ ,
\\
 \lambda^{(\MSb)}(X) &=& \lambda^{(\MSb)}(X= 0) e^{-c_H X} \nonumber \\
&&-(1-e^{-c_H X}) 
\left(
\frac{9}{32\pi^2}y_t^{(\MSb)\,4}
-\frac{15}{512\pi^2}g^{(\MSb)\,4}
-\frac{5}{256\pi^2}g^{(\MSb)\,2}g^{\prime(\MSb)\,2}
\right.
\nonumber\\
&&
\label{eq:awful2}
\hspace{2cm}\left. 
 -\frac{5}{512\pi^2}g^{\prime(\MSb)\,2}g^{\prime(\MSb)\,2}
  -\frac{9}{8\pi^2}\lambda^{(\MSb)\,2}(X=0)
\right)+
\cdots\ .
\end{eqnarray}
Here, the ellipses denote the terms required to cancel the $Q$-dependence of the Higgs VEV through higher order contributions.
As we repeatedly argued in this paper, such an additional requirements are not justified by the de Sitter
swampland conjecture nor by the anthropic principle, though.

In Eq.\,(\ref{eq:Veff1}), the terms which logarithmically depend on $|H|^2$, on the other hand, cannot be eliminated by local counter terms.
Thus, even the meticulously-tuned renormalization conditions in Eqs.\,(\ref{eq:awful1}) and (\ref{eq:awful2}) do not 
cancel the $Q$-dependence of the Higgs VEV completely.
Numerically, we find that these renormalization conditions lead to
\begin{eqnarray}
\left.\frac{d\ln v_H^{(\MSb)} }{dX} \right|_{X=0}\simeq 0.0037 \times c_H\ ,
\end{eqnarray}
at the one-loop level.
Correspondingly, the constraints on the long-range force and the time-varying proton-to-electron mass ratio lead to 
\begin{eqnarray}
c_H \lesssim 0.6 \times 10^{-2}\ ,
\end{eqnarray}
and 
\begin{eqnarray}
c_Q|c_H| \lesssim 0.5\times 10^{-3}\ .
\end{eqnarray}
Therefore, even highly conspiratorial renormalization conditions are still in tension with the de Sitter swampland conjecture.

\subsection{Another constraint}
\begin{figure}[t]
\begin{center}
  \includegraphics[height=0.175\textheight]{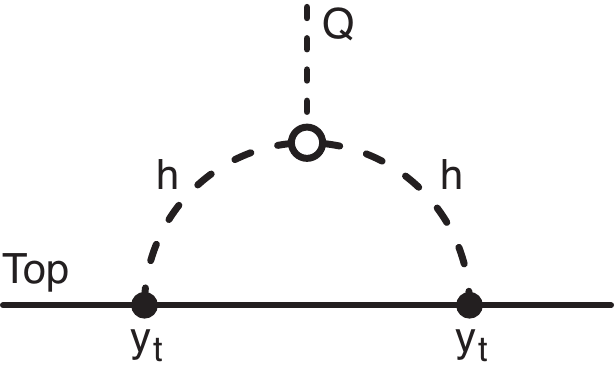}
  \caption{\sl Feynman diagram which radiatively generates $Q\bar{t}t$
    vertex.  Here, $y_t$ denotes the top-quark Yukawa coupling
    constant.}
  \label{fig:qtt}
\end{center}
\end{figure}

So far, we have discussed constraints which are in association with
the $Q$-dependence of the Higgs VEV.  Here, we comment on another
constraint.  It is less severe compared to the previous ones if the
Higgs VEV has $Q$-dependence, but is applicable even if the Higgs VEV
is independent of $Q$ as we show in the following.

In the model of our interest, the coupling of $Q$ to the top quark is
radiatively generated as we have mentioned earlier.  In
Fig.\,\ref{fig:qtt}, we show the Feynman diagram generating the
$Q\bar{t}t$ vertex, where the trilinear scalar interaction shows up by
expanding the scalar potential around $h\simeq v_H$.
As a result, the $Q\bar t t$ vertex is given as%
\footnote{Here, we use the tree-level relation,
  $M_t^{(\MSb)} = y_{t}^{(\MSb)} v_{H}^{(\MSb)}/\sqrt{2}$, which is
  enough at the one-loop level.  }
\begin{eqnarray}
\label{eq:Qtt}
{\cal L}_{\rm eff} \simeq - \frac{1}{16\pi^2} y_t^{(\MSb)\,2}I(M_t^{(\MSb)2}/ 2 M_H^{2(\MSb)})\frac{M_t^{(\MSb)}}{M_{\rm PL}} c_HQ \bar{t}{t}
\simeq -0.003\times \frac{M_t^{(\MSb)}}{M_{\rm PL}} c_HQ \bar{t}{t}\ ,
\end{eqnarray}
where
\begin{eqnarray}
  I(x) = \frac{1-x + x\log\,x}{(1-x)^2} \ .
\end{eqnarray}
Here again, we assume $M_H^{2(\MSb)}(X)\propto e^{-c_HX}$, while the top Yukawa coupling $y_t^{(\MSb)}$ is independent of $X$ as in the 
case of the LQ prescription.
The coupling of $Q$ to the lighter fermions are suppressed by the Yukawa coupling, and hence, less important.

The effective vertex in Eq.\,(\ref{eq:Qtt}) eventually leads to the
coupling to the nucleons,
\begin{eqnarray}
{\cal L}_{\rm eff} \simeq -\frac{2}{27}\left(1- \sum_{q=u,d,s} f^N_q\right)\times 0.003 \times \frac{m_N}{M_{\rm PL}} c_H Q \bar\psi_N\psi_N 
\simeq 2\times 10^{-4} \times\frac{m_N}{M_{\rm PL}} c_H Q \bar\psi_N\psi_N\ .
\nonumber \\
\end{eqnarray}
Thus, the tests on the long range force in Eq.\,(\ref{eq:long}) put a
constraint,
\begin{eqnarray}
  \label{eq:cHinW_noVEVshift}
  |c_H| \lesssim 0.04\ ,
\end{eqnarray}
while the tests of the time-variation of the proton-to-electron mass ratio in Eq.\,\eqref{eq:test} lead to
\begin{eqnarray}
  \label{eq:cHcQinW_noVEVshift}
  c_Q|c_H| \lesssim 0.8\times 10^{-2}\ .
\end{eqnarray}
Those constraints are independent of the ones derived from the
$Q$-dependence of the Higgs VEV.

\section{Conclusions and Discussions}

The recently proposed de Sitter swampland conjecture excludes local
extrema of a scalar potential with a positive energy density in a low
energy effective theory.  Combining with the habitable conditions of
the vacua in the string landscape, the Higgs potential is required to
be retrofitted to have non-trivial couplings to the quintessence field
$Q$ so that the vacuum energy stays very low in the course of
cosmological evolution.

In this paper, we found that the retrofitted Higgs potential
generically predicts that the Higgs VEV becomes dependent on the
amplitude of the quintessence field.  We first discussed that the
Higgs VEV shows a sizable $Q$-dependence based on the general Higgs
potential (see e.g. $V_H^{(b)}$ in Eq.\,(\ref{eq:Higgs3})), which is
consistent with the de Sitter swampland conjecture as well as the
anthropic principle.  We also argued that the overall coupling of the
quintessence filed to the Higgs potential at a high energy scale (see
e.g. $V_H^{(c)}$ in Eq.\,(\ref{eq:Higgs4})) results in a $Q$-dependent
Higgs VEV due to the renormalization-group runnings.  Furthermore, we
also found that, even if $Q$ has the overall coupling to the Higgs
potential at a low energy scale, the Higgs VEV is still $Q$-dependent.
Those conclusions do not depend on the details of the
quintessence-Higgs couplings nor the potential of the quintessence
field as long as they satisfy the de Sitter swampland conjecture and
the anthropic principle.  As a result, we conclude that most of the habitable
vacua with a Higgs potential which satisfies the de Sitter swampland
conjecture predicts sizable $Q$-dependence of the Higgs
VEV unless there is a fundamental reason to exclude a $Q$-dependent
Higgs expectation value.  As we have discussed, the scenario with the
$Q$-dependent Higgs VEV contradicts with the tests of the equivalence
principle as well as the tests of the time-varying proton-to-electron
mass ratio.

Similarly, if there exists a scalar field which provides masses to
colored particles, then the $Q$-dependence of its VEV is required to
be weak enough to avoid the constraints from the long-range force and
the time-varying proton-to-electron mass ratio.  The examples of such
scalar fields include the field which breaks the Peccei-Quinn symmetry
or the Grand Unified gauge symmetry.

In summary, if a quintessence field $Q$ is coupled to the Higgs
potential (as well as to dark energy) to satisfy the swampland
conjecture, the scenario is severely constrained by the
long-range force and the time-dependence of the proton-to-electron
mass ratio.
Unless there exists any additional principle to avoid
these constraints, it seems difficult to find ourselves living in a
vacuum consistent with phenomenological constraints.

\begin{acknowledgments}
  The authors would like to thank M.~Yamazaki and T.~T.~Yanagida for valuable discussion.
  This work is supported in part by Grants-in-Aid for Scientific
  Research from the Ministry of Education, Culture, Sports, Science,
  and Technology (MEXT) KAKENHI, Japan, No.\,15H05889, No.\,16H03991,
  No.\,17H02878, and No.\,18H05542 (M.I.), 
  No.\,16H02189, No.\,26104001, and No.\,26104009  (K.H.), 
  No.\ 16H06490, and No.\ 18K03608 (T.M.), 
  and by the World Premier International Research Center Initiative
  (WPI), MEXT, Japan.
\end{acknowledgments}

\appendix
\section{Time evolution of quintessence field} 
\label{sec:qmotion}
In this Appendix, we briefly discuss how the (real-valued)
quintessence field, $Q$, evolves from the time of the redshift
parameter $z= {\cal O}(1)$, to the present.  (For more detailed
analysis of the evolution of the quintessence field with the potential
in Eq.\,(\ref{eq:quintess}), see
e.g. Refs.~\cite{Copeland:1997et,Ferreira:1997hj,Tsujikawa:2013fta}.)
For $z < {\cal O}(10)$, the Hubble parameter is well approximated by
\begin{eqnarray}
H = H_0 \sqrt{\frac{\Omega_m}{a^3} + \frac{1}{3 H_0^2 M_{\rm PL}^2}\left(\frac{1}{2}\dot{Q}^2+ V_Q(Q)\right)}\ ,
\end{eqnarray}
where the dot denotes the time derivative, and $a$ the scale factor of
the universe.  The first term in the right-hand side denotes the
contribution of the non-relativistic matter (with $\Omega_m \simeq
0.3$ being the density parameter of matter~\cite{Aghanim:2018eyx}),
and the second term the one of the quintessence field $Q$ which plays
a role of the dark energy.  All the other scalar fields than the
quintessence field have settled to their VEVs well before $z \simeq
{\cal O}(1)$.

To demonstrate how $Q$ evolves, we consider the simplest potential of
the quintessence field in Eq.\,(\ref{eq:quintess}).  In the
slow-rolling regime, $c_Q \ll \sqrt{6}$, the energy density of the
quintessence field is dominated by the potential energy.  The equation
of motion of $Q$ is given by
\begin{eqnarray}
\label{eq:EOM}
\ddot{Q}+3 H \dot{Q} = - \q V_Q/\q Q =  3\xi_{cc} H_0^2 M_{\rm PL} c_Q\, e^{-c_Q Q/M_{\rm PL}} \ .
\end{eqnarray}
By using rescaled variables, 
\begin{eqnarray}
X = \frac{Q}{M_{\rm PL}} \ , \quad
x=\sqrt{3} H_0 t \ ,  \quad
X' = \frac{d X}{d x} =  \frac{1}{\sqrt{3}H_0 M_{\rm PL}} \frac{dQ}{dt} \ , \quad
\tilde{V}_Q(X)  =    \xi_{\rm cc} e^{-c_QX }    \ , 
\end{eqnarray}
the Hubble equation and the equation of motion of $Q$ are reduced to
\begin{eqnarray}
\frac{a'}{a} &=& \frac{1}{\sqrt 3}\sqrt{  \frac{\Omega_m}{a^3} +\frac{1}{2}X'{}^2+ \tilde{V}_Q(X)  }\ , \\
X'' &+& \sqrt{3} \sqrt{ \frac{\Omega_m}{a^3} +\frac{1}{2}X'{}^2+ \tilde{V}_Q(X)}   X'= \xi_{cc}  c_Q\, e^{-c_QX}  \ .
\end{eqnarray}
We solve these equations with the boundary conditions,
\begin{eqnarray}
\label{eq:ainit}
a(t_0) &=& 1 \ , \\
\label{eq:Xinit}
X(t_0) & = & 0\ ,\\
\label{eq:Xpinit}
\frac{1}{2}X'{}^2+ \tilde{V}_Q(X)|_{t = t_0} & = & 0.7 \ ,
\end{eqnarray}
with $t_0$ being the present cosmic time.  The initial condition of
$X'$ is taken so that the motion of the quintessence is determined by
the Hubble friction in the matter dominated era.

\begin{figure}[t]
\begin{center}
\begin{minipage}{.47\linewidth}
  \includegraphics[width=\linewidth]{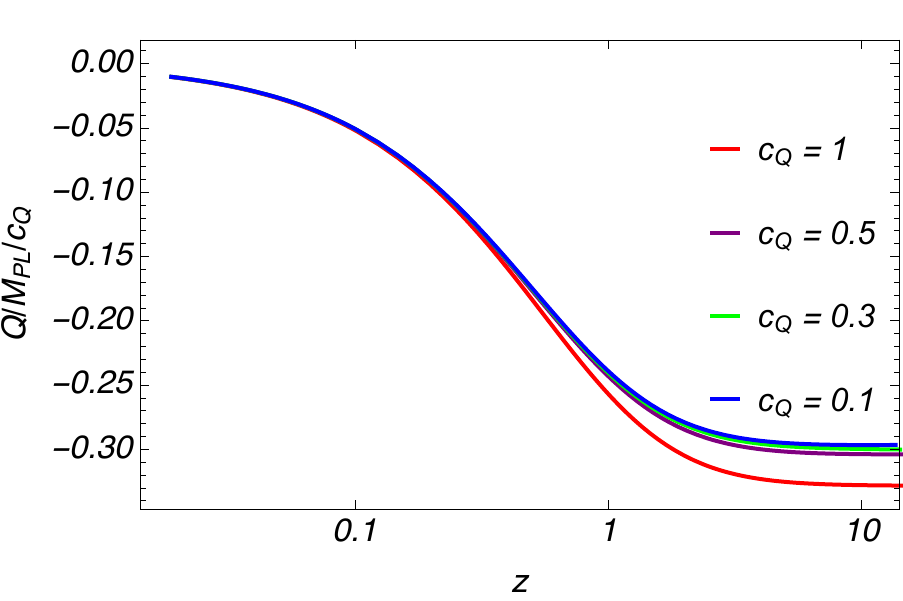}
 \end{minipage}
 \hspace{.6cm}
 \begin{minipage}{.47\linewidth}
  \includegraphics[width=\linewidth]{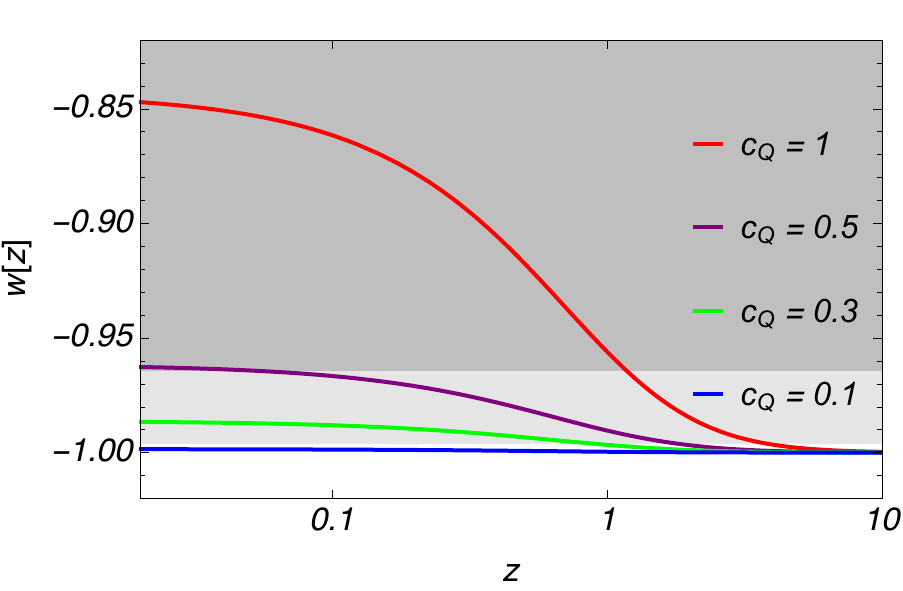}
 \end{minipage}
  \end{center}
\caption{\sl \small  
(Left) The excursion of the quintessence field from $Q(z=0) = 0$ (normalized by $M_{\rm PL} c_Q$) as a function of the redshift parameter $z$ for $c_Q = 1$, $0.5$, $0.3$, and $0.1$.
(Right)  The equation of state of the dark energy as a function of the redshift parameter.
The (light-)gray shaded region is disfavored by the $2\sigma$ ($1\sigma$) limit 
based on CMB, SNe and BAO measurements~\cite{Aghanim:2018eyx},
$w = -1.028 \pm 0.032\,(1\sigma)$.
Here, the energy density of the present universe are taken to be  $\Omega_m = 0.3$
and $\Omega_{DE}= 0.7$.
}
\label{fig:Q}
\end{figure}

In Fig.\,\ref{fig:Q}, we show the evolution of the quintessence field as a function of the redshift parameter $z$ (the left panel).
The figure shows that the field excursion from $z=1$ to $z=0$ is 
\begin{eqnarray}
\label{eq:quintZ}
\left. \frac{{\mit \Delta}Q} {M_{\rm PL}} \right|_{z=1} \simeq - 0.24 \times c_Q  \ , 
\end{eqnarray}
for $c_Q\lesssim 0.5$.
In the right panel, we also show the equation of state of the dark energy,
\begin{eqnarray}
w = -1 + \frac{\dot Q^2}{\frac{1}{2}\dot Q^2 + V_Q(Q)}\ .
\end{eqnarray}
The equation of state is larger than $-1$ as the quintessence field
is in motion.  The (light-)gray shaded region is disfavored by the
$2\sigma$ ($1\sigma$) limit from CMB, SNe and BAO
measurements~\cite{Aghanim:2018eyx}, i.e. $w = -1.028 \pm
0.032\,(1\sigma)$.%
\footnote{This constraint is obtained by assuming that $w$ is constant
  in time.  For more detailed analysis with the evolution of $w$ is
  taken into account, see e.g. Ref.~\cite{Agrawal:2018own}.}  The
figure shows that $c_Q =1$ is excluded while $c_Q < 0.5$ is within the
allowed region.

\bibliographystyle{JHEP}
\bibliography{draft_arxiv.bbl}

\providecommand{\href}[2]{#2}\begingroup\raggedright\begin{thebibliography}{10}

\bibitem{Vafa:2005ui}
C.~Vafa, \emph{{The String landscape and the swampland}},
  \href{https://arxiv.org/abs/hep-th/0509212}{{\ttfamily hep-th/0509212}}.

\bibitem{Ooguri:2006in}
H.~Ooguri and C.~Vafa, \emph{{On the Geometry of the String Landscape and the
  Swampland}},
  \href{https://doi.org/10.1016/j.nuclphysb.2006.10.033}{\emph{Nucl. Phys.}
  {\bfseries B766} (2007) 21--33},
  [\href{https://arxiv.org/abs/hep-th/0605264}{{\ttfamily hep-th/0605264}}].

\bibitem{Ooguri:2016pdq}
H.~Ooguri and C.~Vafa, \emph{{Non-supersymmetric AdS and the Swampland}},
  \href{https://doi.org/10.4310/ATMP.2017.v21.n7.a8}{\emph{Adv. Theor. Math.
  Phys.} {\bfseries 21} (2017) 1787--1801},
  [\href{https://arxiv.org/abs/1610.01533}{{\ttfamily 1610.01533}}].

\bibitem{Obied:2018sgi}
G.~Obied, H.~Ooguri, L.~Spodyneiko and C.~Vafa, \emph{{De Sitter Space and the
  Swampland}},  \href{https://arxiv.org/abs/1806.08362}{{\ttfamily
  1806.08362}}.

\bibitem{Vagnozzi:2018jhn}
S.~Vagnozzi, S.~Dhawan, M.~Gerbino, K.~Freese, A.~Goobar and O.~Mena,
  \emph{{Constraints on the sum of the neutrino masses in dynamical dark energy
  models with $w(z) \geq -1$ are tighter than those obtained in $\Lambda$CDM}},
   \href{https://arxiv.org/abs/1801.08553}{{\ttfamily 1801.08553}}.

\bibitem{Agrawal:2018own}
P.~Agrawal, G.~Obied, P.~J. Steinhardt and C.~Vafa, \emph{{On the Cosmological
  Implications of the String Swampland}},
  \href{https://doi.org/10.1016/j.physletb.2018.07.040}{\emph{Phys. Lett.}
  {\bfseries B784} (2018) 271--276},
  [\href{https://arxiv.org/abs/1806.09718}{{\ttfamily 1806.09718}}].

\bibitem{Andriot:2018wzk}
D.~Andriot, \emph{{On the de Sitter swampland criterion}},
  \href{https://doi.org/10.1016/j.physletb.2018.09.022}{\emph{Phys. Lett.}
  {\bfseries B785} (2018) 570--573},
  [\href{https://arxiv.org/abs/1806.10999}{{\ttfamily 1806.10999}}].

\bibitem{Achucarro:2018vey}
A.~Ach{\'u}carro and G.~A. Palma, \emph{{The string swampland constraints
  require multi-field inflation}},
  \href{https://arxiv.org/abs/1807.04390}{{\ttfamily 1807.04390}}.

\bibitem{Garg:2018reu}
S.~K. Garg and C.~Krishnan, \emph{{Bounds on Slow Roll and the de Sitter
  Swampland}},  \href{https://arxiv.org/abs/1807.05193}{{\ttfamily
  1807.05193}}.

\bibitem{Lehners:2018vgi}
J.-L. Lehners, \emph{{Small-Field and Scale-Free: Inflation and Ekpyrosis at
  their Extremes}},  \href{https://arxiv.org/abs/1807.05240}{{\ttfamily
  1807.05240}}.

\bibitem{Kehagias:2018uem}
A.~Kehagias and A.~Riotto, \emph{{A note on Inflation and the Swampland}},
  \href{https://arxiv.org/abs/1807.05445}{{\ttfamily 1807.05445}}.

\bibitem{Dias:2018ngv}
M.~Dias, J.~Frazer, A.~Retolaza and A.~Westphal, \emph{{Primordial
  Gravitational Waves and the Swampland}},
  \href{https://arxiv.org/abs/1807.06579}{{\ttfamily 1807.06579}}.

\bibitem{Denef:2018etk}
F.~Denef, A.~Hebecker and T.~Wrase, \emph{{The dS swampland conjecture and the
  Higgs potential}},  \href{https://arxiv.org/abs/1807.06581}{{\ttfamily
  1807.06581}}.

\bibitem{Colgain:2018wgk}
E.~{\'O}. Colg{\'a}in, M.~H. P.~M. Van~Putten and H.~Yavartanoo, \emph{{$H_0$
  tension and the de Sitter Swampland}},
  \href{https://arxiv.org/abs/1807.07451}{{\ttfamily 1807.07451}}.

\bibitem{Brandenberger:2018fdd}
R.~Brandenberger, L.~L. Graef, G.~Marozzi and G.~P. Vacca, \emph{{Back-Reaction
  of Super-Hubble Cosmological Perturbations Beyond Perturbation Theory}},
  \href{https://arxiv.org/abs/1807.07494}{{\ttfamily 1807.07494}}.

\bibitem{Paban:2018ole}
S.~Paban and R.~Rosati, \emph{{Inflation in Multi-field Modified DBM
  Potentials}},  \href{https://arxiv.org/abs/1807.07654}{{\ttfamily
  1807.07654}}.

\bibitem{Ghalee:2018qeo}
A.~Ghalee, \emph{{Condensation of a scalar field non-minimally coupled to
  gravity in a cosmological context}},
  \href{https://arxiv.org/abs/1807.08620}{{\ttfamily 1807.08620}}.

\bibitem{Matsui:2018bsy}
H.~Matsui and F.~Takahashi, \emph{{Eternal Inflation and Swampland
  Conjectures}},  \href{https://arxiv.org/abs/1807.11938}{{\ttfamily
  1807.11938}}.

\bibitem{Ben-Dayan:2018mhe}
I.~Ben-Dayan, \emph{{Draining the Swampland}},
  \href{https://arxiv.org/abs/1808.01615}{{\ttfamily 1808.01615}}.

\bibitem{Chiang:2018jdg}
C.-I. Chiang and H.~Murayama, \emph{{Building Supergravity Quintessence
  Model}},  \href{https://arxiv.org/abs/1808.02279}{{\ttfamily 1808.02279}}.

\bibitem{Heisenberg:2018yae}
L.~Heisenberg, M.~Bartelmann, R.~Brandenberger and A.~Refregier, \emph{{Dark
  Energy in the Swampland}},
  \href{https://arxiv.org/abs/1808.02877}{{\ttfamily 1808.02877}}.

\bibitem{Damian:2018tlf}
C.~Damian and O.~Loaiza-Brito, \emph{{Two-field axion inflation and the
  swampland constraint in the flux-scaling scenario}},
  \href{https://arxiv.org/abs/1808.03397}{{\ttfamily 1808.03397}}.

\bibitem{Kinney:2018nny}
W.~H. Kinney, S.~Vagnozzi and L.~Visinelli, \emph{{The Zoo Plot Meets the
  Swampland: Mutual (In)Consistency of Single-Field Inflation, String
  Conjectures, and Cosmological Data}},
  \href{https://arxiv.org/abs/1808.06424}{{\ttfamily 1808.06424}}.

\bibitem{Cicoli:2018kdo}
M.~Cicoli, S.~De~Alwis, A.~Maharana, F.~Muia and F.~Quevedo, \emph{{De Sitter
  vs Quintessence in String Theory}},
  \href{https://arxiv.org/abs/1808.08967}{{\ttfamily 1808.08967}}.

\bibitem{Akrami:2018ylq}
Y.~Akrami, R.~Kallosh, A.~Linde and V.~Vardanyan, \emph{{The landscape, the
  swampland and the era of precision cosmology}},
  \href{https://arxiv.org/abs/1808.09440}{{\ttfamily 1808.09440}}.

\bibitem{Heisenberg:2018rdu}
L.~Heisenberg, M.~Bartelmann, R.~Brandenberger and A.~Refregier, \emph{{Dark
  Energy in the Swampland II}},
  \href{https://arxiv.org/abs/1809.00154}{{\ttfamily 1809.00154}}.

\bibitem{Murayama:2018lie}
H.~Murayama, M.~Yamazaki and T.~T. Yanagida, \emph{{Do We Live in the
  Swampland?}},  \href{https://arxiv.org/abs/1809.00478}{{\ttfamily
  1809.00478}}.

\bibitem{Marsh:2018kub}
M.~C.~D. Marsh, \emph{{The Swampland, Quintessence and the Vacuum Energy}},
  \href{https://arxiv.org/abs/1809.00726}{{\ttfamily 1809.00726}}.

\bibitem{Brahma:2018hrd}
S.~Brahma and M.~Wali~Hossain, \emph{{Avoiding the string swampland in
  single-field inflation: Excited initial states}},
  \href{https://arxiv.org/abs/1809.01277}{{\ttfamily 1809.01277}}.

\bibitem{Choi:2018rze}
K.~Choi, D.~Chway and C.~S. Shin, \emph{{The dS swampland conjecture with the
  electroweak symmetry and QCD chiral symmetry breaking}},
  \href{https://arxiv.org/abs/1809.01475}{{\ttfamily 1809.01475}}.

\bibitem{Das:2018hqy}
S.~Das, \emph{{A note on Single-field Inflation and the Swampland Criteria}},
  \href{https://arxiv.org/abs/1809.03962}{{\ttfamily 1809.03962}}.

\bibitem{Danielsson:2018qpa}
U.~H. Danielsson, \emph{{The quantum swampland}},
  \href{https://arxiv.org/abs/1809.04512}{{\ttfamily 1809.04512}}.

\bibitem{Wang:2018duq}
D.~Wang, \emph{{The multi-feature universe: large parameter space cosmology and
  the swampland}},  \href{https://arxiv.org/abs/1809.04854}{{\ttfamily
  1809.04854}}.

\bibitem{Brandenberger:2018wbg}
R.~H. Brandenberger, \emph{{Beyond Standard Inflationary Cosmology}},
  \href{https://arxiv.org/abs/1809.04926}{{\ttfamily 1809.04926}}.

\bibitem{Han:2018yrk}
C.~Han, S.~Pi and M.~Sasaki, \emph{{Quintessence Saves Higgs Instability}},
  \href{https://arxiv.org/abs/1809.05507}{{\ttfamily 1809.05507}}.

\bibitem{Brandenberger:2018xnf}
R.~Brandenberger, R.~R. Cuzinatto, J.~Fr{\"o}hlich and R.~Namba, \emph{{New
  Scalar Field Quartessence}},
  \href{https://arxiv.org/abs/1809.07409}{{\ttfamily 1809.07409}}.

\bibitem{Matsui:2018xwa}
H.~Matsui, F.~Takahashi and M.~Yamada, \emph{{Isocurvature Perturbations of
  Dark Energy and Dark Matter from the Swampland Conjecture}},
  \href{https://arxiv.org/abs/1809.07286}{{\ttfamily 1809.07286}}.

\bibitem{Ratra:1987rm}
B.~Ratra and P.~J.~E. Peebles, \emph{{Cosmological Consequences of a Rolling
  Homogeneous Scalar Field}},
  \href{https://doi.org/10.1103/PhysRevD.37.3406}{\emph{Phys. Rev.} {\bfseries
  D37} (1988) 3406}.

\bibitem{Wetterich:1987fm}
C.~Wetterich, \emph{{Cosmology and the Fate of Dilatation Symmetry}},
  \href{https://doi.org/10.1016/0550-3213(88)90193-9}{\emph{Nucl. Phys.}
  {\bfseries B302} (1988) 668--696},
  [\href{https://arxiv.org/abs/1711.03844}{{\ttfamily 1711.03844}}].

\bibitem{Tsujikawa:2013fta}
S.~Tsujikawa, \emph{{Quintessence: A Review}},
  \href{https://doi.org/10.1088/0264-9381/30/21/214003}{\emph{Class. Quant.
  Grav.} {\bfseries 30} (2013) 214003},
  [\href{https://arxiv.org/abs/1304.1961}{{\ttfamily 1304.1961}}].

\bibitem{Wagner:2012ui}
T.~A. Wagner, S.~Schlamminger, J.~H. Gundlach and E.~G. Adelberger,
  \emph{{Torsion-balance tests of the weak equivalence principle}},
  \href{https://doi.org/10.1088/0264-9381/29/18/184002}{\emph{Class. Quant.
  Grav.} {\bfseries 29} (2012) 184002},
  [\href{https://arxiv.org/abs/1207.2442}{{\ttfamily 1207.2442}}].

\bibitem{Martins:2017yxk}
C.~J. A.~P. Martins, \emph{{The status of varying constants: a review of the
  physics, searches and implications}},
  \href{https://arxiv.org/abs/1709.02923}{{\ttfamily 1709.02923}}.

\bibitem{Aghanim:2018eyx}
{\scshape Planck} collaboration, N.~Aghanim et~al., \emph{{Planck 2018 results.
  VI. Cosmological parameters}},
  \href{https://arxiv.org/abs/1807.06209}{{\ttfamily 1807.06209}}.

\bibitem{Bousso:2000xa}
R.~Bousso and J.~Polchinski, \emph{{Quantization of four form fluxes and
  dynamical neutralization of the cosmological constant}},
  \href{https://doi.org/10.1088/1126-6708/2000/06/006}{\emph{JHEP} {\bfseries
  06} (2000) 006}, [\href{https://arxiv.org/abs/hep-th/0004134}{{\ttfamily
  hep-th/0004134}}].

\bibitem{Susskind:2003kw}
L.~Susskind, \emph{{The Anthropic landscape of string theory}},
  \href{https://arxiv.org/abs/hep-th/0302219}{{\ttfamily hep-th/0302219}}.

\bibitem{Tegmark:2005dy}
M.~Tegmark, A.~Aguirre, M.~Rees and F.~Wilczek, \emph{{Dimensionless constants,
  cosmology and other dark matters}},
  \href{https://doi.org/10.1103/PhysRevD.73.023505}{\emph{Phys. Rev.}
  {\bfseries D73} (2006) 023505},
  [\href{https://arxiv.org/abs/astro-ph/0511774}{{\ttfamily
  astro-ph/0511774}}].

\bibitem{Shifman:1978zn}
M.~A. Shifman, A.~I. Vainshtein and V.~I. Zakharov, \emph{{Remarks on Higgs
  Boson Interactions with Nucleons}},
  \href{https://doi.org/10.1016/0370-2693(78)90481-1}{\emph{Phys. Lett.}
  {\bfseries 78B} (1978) 443--446}.

\bibitem{Hoferichter:2017olk}
M.~Hoferichter, P.~Klos, J.~Men{\'e}ndez and A.~Schwenk, \emph{{Improved limits
  for Higgs-portal dark matter from LHC searches}},
  \href{https://doi.org/10.1103/PhysRevLett.119.181803}{\emph{Phys. Rev. Lett.}
  {\bfseries 119} (2017) 181803},
  [\href{https://arxiv.org/abs/1708.02245}{{\ttfamily 1708.02245}}].

\bibitem{Crivellin:2013ipa}
A.~Crivellin, M.~Hoferichter and M.~Procura, \emph{{Accurate evaluation of
  hadronic uncertainties in spin-independent WIMP-nucleon scattering:
  Disentangling two- and three-flavor effects}},
  \href{https://doi.org/10.1103/PhysRevD.89.054021}{\emph{Phys. Rev.}
  {\bfseries D89} (2014) 054021},
  [\href{https://arxiv.org/abs/1312.4951}{{\ttfamily 1312.4951}}].

\bibitem{Hall:2014dfa}
L.~J. Hall, D.~Pinner and J.~T. Ruderman, \emph{{The Weak Scale from BBN}},
  \href{https://doi.org/10.1007/JHEP12(2014)134}{\emph{JHEP} {\bfseries 12}
  (2014) 134}, [\href{https://arxiv.org/abs/1409.0551}{{\ttfamily 1409.0551}}].

\bibitem{Harnik:2006vj}
R.~Harnik, G.~D. Kribs and G.~Perez, \emph{{A Universe without weak
  interactions}}, \href{https://doi.org/10.1103/PhysRevD.74.035006}{\emph{Phys.
  Rev.} {\bfseries D74} (2006) 035006},
  [\href{https://arxiv.org/abs/hep-ph/0604027}{{\ttfamily hep-ph/0604027}}].

\bibitem{Campbell:1994bf}
B.~A. Campbell and K.~A. Olive, \emph{{Nucleosynthesis and the time dependence
  of fundamental couplings}},
  \href{https://doi.org/10.1016/0370-2693(94)01652-S}{\emph{Phys. Lett.}
  {\bfseries B345} (1995) 429--434},
  [\href{https://arxiv.org/abs/hep-ph/9411272}{{\ttfamily hep-ph/9411272}}].

\bibitem{Weinberg:1987vp}
E.~J. Weinberg and A.-q. Wu, \emph{{UNDERSTANDING COMPLEX PERTURBATIVE
  EFFECTIVE POTENTIALS}},
  \href{https://doi.org/10.1103/PhysRevD.36.2474}{\emph{Phys. Rev.} {\bfseries
  D36} (1987) 2474}.

\bibitem{Ford:1992pn}
C.~Ford, I.~Jack and D.~R.~T. Jones, \emph{{The Standard model effective
  potential at two loops}}, \href{https://doi.org/10.1016/0550-3213(92)90165-8,
  10.1016/S0550-3213(97)00532-4}{\emph{Nucl. Phys.} {\bfseries B387} (1992)
  373--390}, [\href{https://arxiv.org/abs/hep-ph/0111190}{{\ttfamily
  hep-ph/0111190}}].

\bibitem{Martin:2013gka}
S.~P. Martin, \emph{{Three-loop Standard Model effective potential at leading
  order in strong and top Yukawa couplings}},
  \href{https://doi.org/10.1103/PhysRevD.89.013003}{\emph{Phys. Rev.}
  {\bfseries D89} (2014) 013003},
  [\href{https://arxiv.org/abs/1310.7553}{{\ttfamily 1310.7553}}].

\bibitem{Degrassi:2012ry}
G.~Degrassi, S.~Di~Vita, J.~Elias-Miro, J.~R. Espinosa, G.~F. Giudice,
  G.~Isidori et~al., \emph{{Higgs mass and vacuum stability in the Standard
  Model at NNLO}}, \href{https://doi.org/10.1007/JHEP08(2012)098}{\emph{JHEP}
  {\bfseries 08} (2012) 098},
  [\href{https://arxiv.org/abs/1205.6497}{{\ttfamily 1205.6497}}].

\bibitem{Tanabashi:2018oca}
{\scshape Particle Data Group} collaboration, M.~Tanabashi et~al.,
  \emph{{Review of Particle Physics}},
  \href{https://doi.org/10.1103/PhysRevD.98.030001}{\emph{Phys. Rev.}
  {\bfseries D98} (2018) 030001}.

\bibitem{Martin:2014cxa}
S.~P. Martin and D.~G. Robertson, \emph{{Higgs boson mass in the Standard Model
  at two-loop order and beyond}},
  \href{https://doi.org/10.1103/PhysRevD.90.073010}{\emph{Phys. Rev.}
  {\bfseries D90} (2014) 073010},
  [\href{https://arxiv.org/abs/1407.4336}{{\ttfamily 1407.4336}}].

\bibitem{Martin:2005qm}
S.~P. Martin and D.~G. Robertson, \emph{{TSIL: A Program for the calculation of
  two-loop self-energy integrals}},
  \href{https://doi.org/10.1016/j.cpc.2005.08.005}{\emph{Comput. Phys. Commun.}
  {\bfseries 174} (2006) 133--151},
  [\href{https://arxiv.org/abs/hep-ph/0501132}{{\ttfamily hep-ph/0501132}}].

\bibitem{Copeland:1997et}
E.~J. Copeland, A.~R. Liddle and D.~Wands, \emph{{Exponential potentials and
  cosmological scaling solutions}},
  \href{https://doi.org/10.1103/PhysRevD.57.4686}{\emph{Phys. Rev.} {\bfseries
  D57} (1998) 4686--4690},
  [\href{https://arxiv.org/abs/gr-qc/9711068}{{\ttfamily gr-qc/9711068}}].

\bibitem{Ferreira:1997hj}
P.~G. Ferreira and M.~Joyce, \emph{{Cosmology with a primordial scaling
  field}}, \href{https://doi.org/10.1103/PhysRevD.58.023503}{\emph{Phys. Rev.}
  {\bfseries D58} (1998) 023503},
  [\href{https://arxiv.org/abs/astro-ph/9711102}{{\ttfamily
  astro-ph/9711102}}].

\end{thebibliography}\endgroup

\end{document}